\documentclass[12pt,a4paper]{article}
\usepackage{a4wide}
\usepackage{amsmath}
\usepackage{amssymb}
\usepackage{epsfig}
\usepackage{subfigure}
\usepackage{exscale}
\usepackage{float}
\usepackage{bbm}
\usepackage[numbers,sort&compress]{natbib}
\usepackage{pst-plot, pstricks,pst-math}
\usepackage{fancybox,amssymb,color}
\usepackage{graphicx}
\usepackage{pstricks, color, graphicx, epsfig, psfrag}
\usepackage{amsfonts,amsmath,amssymb,slashed}
\usepackage{dsfont}
\usepackage{bbm,bm}
\usepackage{fancyhdr, a4wide}
\usepackage[english]{babel}
\usepackage{subfigure}

\makeatletter
\@addtoreset{equation}{section}
\makeatother

\title{Thermodynamics of Ferromagnetic Spin Chains in a Magnetic Field: Impact of the Spin-Wave Interaction}

\author{Christoph P.\ Hofmann$^a$ \\ \\
\normalsize {$^a$ Facultad de Ciencias, Universidad de Colima} \\
\vspace{0.3cm}
\normalsize {Bernal D\'iaz del Castillo 340, Colima C.P.\ 28045, Mexico} \\}

\begin{document}

\maketitle

\begin{abstract} \normalsize

\end{abstract}

The thermodynamic properties of ferromagnetic spin chains have been the subject of many publications. Still, the problem of how the
spin-wave interaction manifest itself in these low-temperature series has been neglected. Using the method of effective Lagrangians, we
explicitly evaluate the partition function of ferromagnetic spin chains at low temperatures and in the presence of a magnetic field up to
three loops in the perturbative expansion where the spin-wave interaction sets in. We discuss in detail the renormalization and numerical
evaluation of a particular three-loop graph and derive the low-temperature series for the free energy density, energy density, heat
capacity, entropy density, as well as the magnetization and the susceptibility. In the low-temperature expansion for the free energy
density, the spin-wave interaction starts manifesting itself at order $T^{5/2}$. In the pressure, the coefficient of the $T^{5/2}$-term is
positive, indicating that the spin-wave interaction is repulsive. While it is straightforward to go up to three-loop order in the
effective loop expansion, the analogous calculation on the basis of conventional condensed matter methods, such as spin-wave theory,
appears to be beyond reach.

% \pacs{75.40.Cx, 75.50.Ee, 12.39.Fe, 11.10.Wx} %O.K. (22.02.13)

\maketitle

\section{Introduction}
\label{Intro}

In the present study we rigorously answer the question of how the spin-wave interaction manifests itself in the low-temperature properties
of ferromagnetic spin chains in a weak magnetic field. Using the systematic method of effective Lagrangians, in the very recent article
\citep{Hof12c}, it was argued that the spin-wave interaction only starts showing up at the three-loop level. However, the explicit
evaluation of the various Feynman graphs contributing at this order to the partition function, has not been addressed in that reference.
This quite elaborate task is the subject of the present article. We then provide the low-temperature series for the free energy density,
energy density, heat capacity, entropy density, as well as the magnetization and the susceptibility.

The effective Lagrangian method relies on the fact that the low-energy dynamics of the system is captured by the Goldstone bosons, which
result from the spontaneously broken global symmetry. In the present case, the spin rotation symmetry of the Heisenberg ferromagnet is
spontaneously broken, O(3) $\to$ O(2), and the spin-waves or magnons emerge as Goldstone bosons. Conceptually, it is quite remarkable that
the effective Lagrangian method works in one spatial dimension. It is well-known that in a Lorentz-invariant framework, where the
Goldstone bosons (pions, kaons, $\eta$-particle) follow a linear, i.e. relativistic, dispersion relation, the method fails in one spatial
dimension. However, the ferromagnet, where the spin waves obey a quadratic dispersion law, is quite peculiar: here the systematic loop
expansion perfectly works as we explain below.

In the low-temperature expansion of the free energy density, the spin-wave interaction generates a term of order $T^{5/2}$. The general
structure of this series is discussed, and the question of which contributions are due to free magnon particles and which ones are due to
the spin-wave interaction is thoroughly answered. In view of the nonperturbatively generated energy gap, we also critically examine the
range of validity of the effective low-temperature series, pointing out that it is not legitimate to take the limit of a zero magnetic
field.

The thermodynamic properties of ferromagnetic spin chains have attracted a lot of attention over the past few decades and many methods
have been used to study these interesting one-dimensional systems. While early investigations were based on the Bethe ansatz
\citep{Tak71,Tak73,Sch85,TY85,YT86,Sch86,LS87,Yam90,GBT07},
modified spin-wave theory was the method advocated in Refs.~\citep{Tak86,Tak87a,SB06}.
Further methods used to address ferromagnetic spin chains include Schwinger-boson mean-field theory
\citep{AA90a,AA90b},
Green functions \citep{SSI94,KY72,CAEK89,HCHB01,JIRK04,APPC08,JIBJ08,LCSWD11},
variants of spin-wave theory \citep{KSK03},
scaling methods \citep{Kop89,CAEK89,NT94,NHT95a,NHT95b,RS95,TNS96,Sac06},
numerical simulations \citep{CL83,Lyk83,FU86,CCL87,DL90,SSC97,GPL05,JIBJ08}, and yet other approaches
\citep{ZE90,YG99,CTVV00,CCC01,DK06,DK12}. Given this abundant literature on ferromagnetic spin chains, it is really surprising that the
effect of the spin-wave interaction has been largely neglected. In particular, although ferromagnetic spin chains can be solved exactly by
e.g. the Bethe ansatz, the low-temperature series derived from these exact results all refer to either a tiny or a zero magnetic field,
which does not cover the domain we are interested in here.

We emphasize that in the problem under consideration, the effective field theory approach is more efficient than conventional condensed
matter methods such as spin-wave theory, as it allows one to {\it systematically} go to higher orders in the low-temperature expansion --
beyond the results provided in the literature. Above all -- for the first time, to the best of our knowledge -- the manifestation of the
spin-wave interaction in the low-temperature behavior of ferromagnetic spin chains in a magnetic field is discussed in a systematic
manner. Almost all previous theoretical studies that analyzed the structure of the low-temperature series for ferromagnetic spin chains
were restricted to the idealized picture of the free magnon gas. One exception is Ref.~\citep{KSK03} which, however, refers to a tiny
magnetic field and appears to be not quite consistent, as we point out in Sec.~\ref{Thermodyn}.

The rest of the paper is organized as follows. In Sec.~\ref{EFT} we provide the reader with some basic aspects of the effective Lagrangian
technique. The low-temperature expansion of the partition function up to three-loop order is derived in Sec.~\ref{PartitionFunction}. The
nontrivial part concerns the renormalization of a particular three-loop graph which is discussed in detail in Sec.~\ref{Renorm}. The
low-temperature series for the free energy density, pressure, energy density, entropy density, heat capacity, as well as the magnetization
and the susceptibility for ferromagnetic spin chains in a magnetic field are given in Sec.~\ref{Thermodyn}. While our conclusions are
presented in Sec.~\ref{Summary}, details on the numerical evaluation of a specific three-loop graph are discussed in two appendices.

The model-independent and systematic effective Lagrangian method, unfortunately, is still not very well known among condensed matter
physicists. We would like to convince the reader that this method indeed represents an alternative and rigorous theoretical framework
to address condensed matter systems, by providing a list of articles which are also based on this method. Ferromagnets and
antiferromagnets in three and two space dimensions were considered in
Refs.~\citep{Hof02,Hof11a,HL90,HN91,HN93,Leu94a,Hof99a,Hof99b,RS99a,RS99b,RS00,Hof01,Hof10,Hof11b,Hof11c,Hof12a,Hof12b}.
Two-dimensional antiferromagnets, doped with either holes or electrons, which represent the precursors of high-temperature superconductors
were analyzed in Refs.~\citep{KMW05,BKMPW06,BKPW06,BHKPW07,BHKMPW07,BHKPW08,JKHW09,BHKMPW09,KBWHJW12,VHJW12}. Moreover, it was
demonstrated in Refs.~\citep{GHKW10,WJ94,GHJNW09,JW11,GHJPSW11} that the effective Lagrangian technique is perfectly consistent both with
numerical simulations based on the loop-cluster algorithm and an analytically solvable microscopic model in one spatial dimension.

\section{Effective Lagrangian Method}
\label{EFT}

In a very recent article, Ref.~\citep{Hof12c}, the low-temperature expansion of partition function for the ferromagnetic spin chain in a
weak magnetic field was evaluated up to two loops. Here we perform the analysis up to three-loop order, where the spin-wave interaction
comes into play. Essential aspects of the effective Lagrangian method at finite temperature have been discussed in section 2 of
Ref.~\citep{Hof12c} and will not be repeated here in detail. Below, we just focus on some basic ingredients of the method. Although
section 2 of Ref.~\citep{Hof12c} is self-contained and contains all the necessary information to understand the present calculation, the
interested reader may still find more details on finite-temperature effective Lagrangians in appendix A of Ref.~\citep{Hof11a} and in the
various references given therein.

The systematic construction of the effective field theory is based on an inspection of the symmetries inherent in the underlying theory.
In the present case, the effective Lagrangian, or more precisely, the effective action
\begin{equation}
{\cal S}_{eff} = \int d^2 x \, {\cal L}_{eff}
\end{equation}
describing the ferromagnetic spin chain, must share all the symmetries of the underlying Heisenberg model. These include the spontaneously
broken spin rotation symmetry O(3), parity and time reversal. One also has to identify the relevant low-energy degrees of freedom entering
the effective description. In the case of the Heisenberg ferromagnet, these are the two real magnon fields -- or the physical magnon
particle -- that arise due to the spontaneously broken spin symmetry O(3) $\to$ O(2).

The various terms in the effective Lagrangian are organized systematically according to the number of space and time derivatives which act
on the magnon fields. At low energies or temperatures, terms which contain only a few derivatives are the dominant ones, while terms with
a larger number of derivatives are suppressed \citep{Wei79,GL85,Leu94b}. The effective Lagrangian ${\cal L}_{eff}$ thus amounts to a
systematic derivative expansion, or, equivalently, an expansion in powers of energy and momentum. Hence the quantities of physical
interest (partition function, free energy density, magnetization, etc.) derived from  ${\cal L}_{eff}$, also correspond to expansions in
powers of momentum which -- at finite temperature -- translate into expansions in powers of temperature.

The leading-order effective Lagrangian for the one-dimensional ferromagnet is of momentum order $p^2$ and reads
\citep{Leu94a}
\begin{equation}
\label{leadingLagrangian}
{\cal L}^2_{eff} = \Sigma \frac{\epsilon_{ab} {\partial}_0 U^a U^b}{1+ U^3}
+ \Sigma \mu H U^3 - \mbox{$\frac{1}{2}$} F^2 {\partial}_{x_1} \! U^i {\partial}_{x_1} \! U^i \, .
\end{equation}
The fundamental object is the three-dimensional magnetization unit vector $U^i = (U^a, U^3)$, where the two real components $U^a \,(a=1,2)$
describe the spin-wave degrees of freedom. The quantity $H$ is the magnetic field which points into the third direction,
${\vec H} = (0,0,H)$ with $H = |\vec H| > 0$. While the derivative structure of the above terms is determined by the symmetries of the
underlying theory, the two a priori unknown low-energy coupling constants -- the spontaneous magnetization at zero temperature $\Sigma$,
and the constant $F$ -- have to be fixed experimentally, in a numerical simulation or by comparison with the microscopic theory. It is
important to point out that one time derivative (${\partial}_0$) counts as two space derivatives (${\partial}_{x_1} {\partial}_{x_1}$),
i.e., two powers of momentum are on the same footing as one power of energy or temperature: $k^2 \propto \omega, T$. This is
characteristic of ferromagnetic systems where the magnons display a quadratic dispersion relation.

The next-to-leading-order effective Lagrangian for the ferromagnetic spin chain is of order $p^4$ and involves the two effective coupling
constants $l_1$ and $l_3$ \citep{Hof12c},
\begin{equation}
\label{Leff4}
{\cal L}^4_{eff} = l_1 {( {\partial}_{x_1} U^i \, {\partial}_{x_1} U^i )}^2 + l_3 \, {\partial}_{x_1}^2 U^i \, {\partial}_{x_1}^2  U^i \, .
\end{equation}
Higher-order pieces in the effective Lagrangian are not needed for the present calculation.

\begin{figure}
\includegraphics[width=15.5cm]{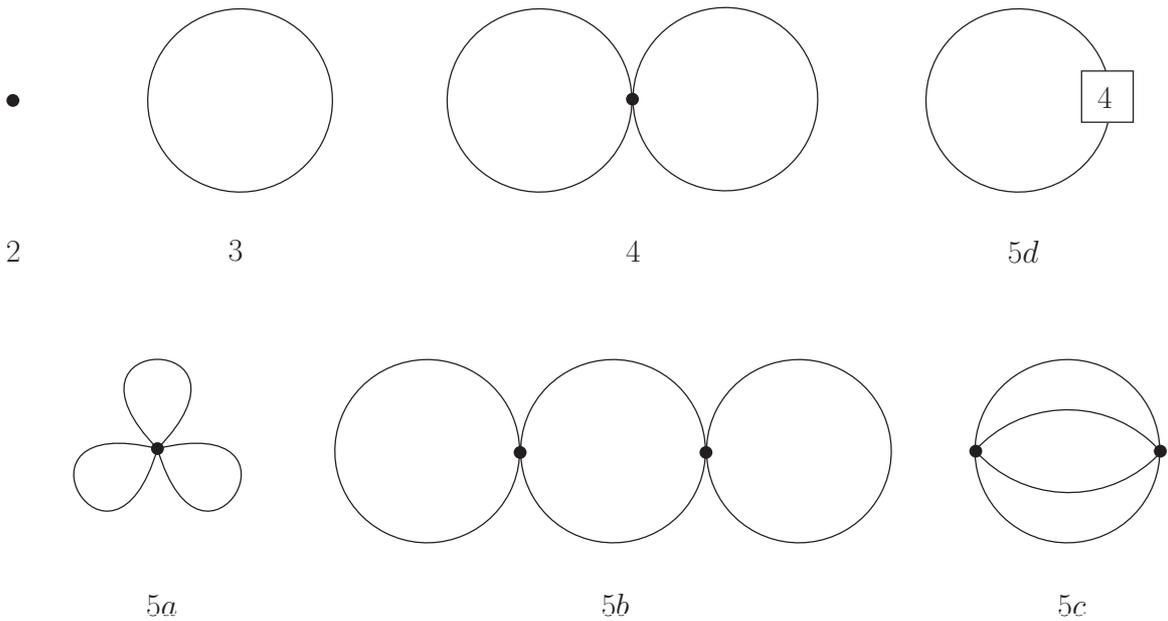}

\caption{Feynman diagrams referring to the low-temperature expansion of the partition function of ferromagnetic spin chains up to and
including order $p^5$. While the number 4 attached to the vertex in diagram 5d corresponds to the next-to-leading-order Lagrangian 
${\cal L}^4_{eff}$, vertices corresponding to the leading term ${\cal L}^2_{eff}$ are denoted by a filled circle. Note that loops are
suppressed by one momentum power.}
\label{figure1}
\end{figure} 

The systematic perturbative evaluation of the partition function relies on the suppression of loops by some power of momentum. In one
spatial dimension, ferromagnetic loops are suppressed by one power of momentum \citep{Hof12c}. The corresponding Feynman graphs for the
partition function up to order $p^5$ are depicted in Fig.~\ref{figure1}. The leading temperature-dependent contribution stems from the
one-loop graph 3 which is of order $p^3$, as it involves a vertex from ${\cal L}^2_{eff}$ ($p^2$) and one loop ($p$). The one-loop diagram
5d with an insertion from ${\cal L}^4_{eff}$ is of order $p^5$, as it involves ${\cal L}^4_{eff}$ ($p^4$) and one loop ($p$). Finally, the
two-loop (three-loop) diagrams are of order $p^4$ ($p^5$) as they involve one (two) more loops with respect to diagram 3. Again, more
details on the perturbative evaluation of the partition function can be found in section 2 of Ref.~\citep{Hof12c}.

\section{Evaluation of the Partition Function up to Three-Loop Order}
\label{PartitionFunction}

The low-temperature expansion of the partition function for the ferromagnetic spin chain in a weak magnetic field was derived in
Ref.~\citep{Hof12c} up to two-loop order. The relevant diagrams 2, 3, 4 and 5d of Fig.~\ref{figure1} lead to the following expression for
the free energy density,
\begin{equation}
\label{FreeCollect}
z = - \Sigma \mu H - \frac{1}{2 \pi^{\frac{1}{2}} \gamma^{\frac{1}{2}}} \, T^{\frac{3}{2}} \, \sum^{\infty}_{n=1}
\frac{e^{- \mu H n \beta}}{n^{\frac{3}{2}}}
- \frac{3 l_3}{4 {\pi}^{\frac{1}{2}} \Sigma {\gamma}^{\frac{5}{2}}} \, T^{\frac{5}{2}} \sum^{\infty}_{n=1}
\frac{e^{- \mu H n \beta}}{n^{\frac{5}{2}}} + {\cal O}(p^5) \, ,
\end{equation}
where $\beta \equiv 1/T$. The contributions of order $T^{3/2}$ and $T^{5/2}$ arise from the one-loop graphs and thus describe
noninteracting spin waves. While the former term only depends on the leading-order effective constants $\Sigma$ and $F$
($\gamma = F^2/\Sigma$, spin stiffness), the latter also involves the next-to-leading-order effective coupling $l_3$ from
${\cal L}^4_{eff}$. It is quite remarkable that the spin-wave interaction does not enter at the two-loop level in the above low-temperature
expansion. This is because the diagram 4 of order $T^2$ is 0 due to parity \citep{Hof12c}. In order to discuss the impact of the
spin-wave interaction, we thus have to go up to three-loop order. Note that the effective coupling constants can be expressed in terms of
microscopic quantities as follows \citep{Hof12c}:
\begin{equation}
\label{micro}
\Sigma = \frac{S}{a} \, , \qquad F^2 = J S^2 a \, , \qquad \gamma = \frac{F^2}{\Sigma} = J S a^2 \, , \qquad l_3 = \frac{J S^2 a^3}{24}
\, .
\end{equation}
Here $a$ is the distance between two spins and $J$ represents the exchange integral of the Heisenberg model. It is important to point out
that the exchange integral $J$ represents the natural scale of the underlying theory. So whenever we talk of low temperature or weak
magnetic field, we mean that the ratios $T/J$ and $H/J$ are small.

For reasons that will become more evident in Sec.~\ref{Renorm}, it is advantageous to use the real-space imaginary-time representation for
the propagators, rather than to work with the momentum-frequency representation. The thermal propagator $G(x)$ amounts to
\begin{equation}
\label{ThermalPropagator}
G(x) \, = \, \sum_{n \,= \, - \infty}^{\infty} \Delta(x_1, x_4 + n \beta) \, ,
\end{equation}
where $\Delta(x)$ is the Euclidean zero temperature propagator referring to ferromagnetic magnons,
\begin{equation}
\label{Propagator}
\Delta (x) \, = \, \int \! \, \frac{d k_4 d k}{(2\pi)^2}
\frac{e^{i k x_1 - i k_4 x_4}}{\gamma k^2 - i k_4 + \mu H} \, . 
\end{equation}
The connection between the momentum-frequency representation and the real-space imaginary-time representation is given by a Fourier
transform. In one spatial dimension, the thermal Matsubara propagator takes the form
\begin{equation}
{\cal G}(k, \omega_n) =   \int_0^{\beta} dx_4 \int \! dx_1  \, e^{i \omega_n x_4 - i k x_1}  \Delta(x_1, x_4) \, ,
\end{equation}
where
\begin{equation}
\omega_n = 2\pi n / \beta \, .
\end{equation}

Moreover, it is convenient to use dimensional regularization in the effective theory, since this regularization scheme respects the
symmetries of the theory. The thermal propagator, regularized in the spatial dimension $d_s$, is then given by
\begin{equation}
\label{ThermProp}
G(x) = \frac{1}{(4 \pi \gamma)^{\frac{d_s}{2}}} \,
\sum^{\infty}_{n \, = \, - \infty} \frac{1}{x_n^{\frac{d_s}{2}}} \,
e^{ - \frac{{\vec x}^2}{4 \gamma x_n} - \mu H x_n} \, \Theta (x_n) \, ,
\end{equation}
with
\begin{equation}
x_n \, \equiv \, x_4 + n \beta \, .
\end{equation}
Finally we use the notation,
\begin{equation}
\label{definitionsThermProp}
G_1 \equiv \Big[ G(x) \Big]_{x=0} \, , \quad
G_{\Delta} \equiv \Big[ {\Delta} G(x) \Big]_{x=0} \, ,
\end{equation}
where $\Delta$ is the Laplace operator in the regularized spatial dimension. It should not be confused with the symbol $\Delta (x)$, which
represents the Euclidean propagator at zero temperature.

The quantities $G_1$ and $G_{\Delta}$ can be decomposed into two parts: a piece that does depend on temperature, and a piece that is
temperature independent,
\begin{equation}
\label{DecompositionPropagator}
G_1 \; = \; G^T_1 \, + \, G^0_1 \, , \qquad G_{\Delta} \; = \; G^T_{\Delta} \, + \, G^0_{\Delta} \, .
\end{equation}
The explicit expressions for $G^T_1$ and $G^T_{\Delta}$, regularized in the parameter $d_s$, read
\begin{eqnarray}
\label{ThermProp(x=0)}
G^T_1 & = & \frac{1}{(4 \pi \gamma)^{\frac{d_s}{2}}} \, \sum^{\infty}_{n=1}
\frac{e^{- \mu H n \beta}}{(n \beta)^{\frac{d_s}{2}}} \, , \nonumber \\
G^T_{\Delta} & = & \frac{1}{(4 \pi \gamma)^{\frac{d_s}{2}}} \,
\Big(\! - \frac{d_s}{2\gamma} \Big) \sum^{\infty}_{n=1} \frac{e^{ - \, \mu H n
\beta}}{(n\beta)^{\frac{d_s}{2} + 1}} \, .
\end{eqnarray}
The temperature-independent pieces $G^0_1$ and $G^0_{\Delta}$ do not contribute to the partition function: in dimensional regularization,
these expressions vanish exactly. Hence, the only contributions which matter in our evaluation are those which are temperature dependent.
These are finite if the regularization is removed, i.e., if the limit $d_s \to 1$ is taken.

After these technical remarks, we now address the three-loop graphs. Note that they only involve vertices from the leading-order
Lagrangian ${\cal L}^2_{eff}$. Graph 5a amounts to a product of three thermal propagators (and space derivatives thereof), which have to be
evaluated at the origin,
\begin{equation}
z_{5a} \, = \, - \frac{F^2}{{\Sigma}^3} \, G_{\Delta} {(G_1)}^2 \, .
\end{equation}
The three-loop graph 5b, remarkably, does not contribute to the partition function,
\begin{equation}
z_{5b} = 0 \, .
\end{equation}
Finally, the cateye graph 5c yields
\begin{equation}
\label{cateye}
z_{5c} \, = \, - \frac{F^4}{2{\Sigma}^4} \, I \, + \, \frac{F^2}{{\Sigma}^3 } G_{\Delta} {(G_1)}^2 \, .
\end{equation}
The quantity $I$ stands for the following integral over the torus,
\begin{equation}
\label{integralI}
I = {\int}_{\! \! \! {\cal T}} \! \! d^{d_s+1}x \, {\partial}_r G \, {\partial}_r G \, {\partial}_s {\tilde G} \, {\partial}_s {\tilde G} \, ,
\qquad r,s = x_1, \dots , x_{d_s} \, ,
\end{equation}
displaying a product of four thermal propagators, where
\begin{equation}
G = G(x) \, , \qquad {\tilde G} = G(-x) \, .
\end{equation}
Since the second term in (\ref{cateye}) cancels the contribution from graph 5a, the only relevant piece at the three-loop level is the one
involving the integral $I$. As this contribution does not just correspond to a product of thermal propagators (or derivatives thereof) to
be evaluated at $x=0$, its renormalization and numerical evaluation is much more involved. These issues are considered in detail in the
following section, as well as in appendices \ref{appendixA} and \ref{appendixB}.

\section{Cateye Graph: Renormalization}
\label{Renorm}

In order to analyze potential ultraviolet divergences in the three-loop graph 5c, it is essential that we use the real-space
imaginary-time representation for the propagators. We adopt the strategy outlined in Ref.~\citep{GL89}, where the same three-loop graph
was discussed within a Lorentz-invariant context.

The relevant integral,
\begin{displaymath}
I \, = {\int}_{\! \! \! {\cal T}} \! \! d^{d_s+1}x \, {\partial}_r G \,
{\partial}_r G \, {\partial}_s {\tilde G} \, {\partial}_s {\tilde G},
\end{displaymath}
exhibits a product of four thermal propagators. Each one of them we split into two parts,
\begin{equation}
\label{decompo}
G(x) = G^T(x) + \Delta(x) \, .
\end{equation}
While the temperature-dependent piece $G^T(x)$ is finite in the limit $d_s \! \to \! 1$,  the zero-temperature propagator $\Delta(x)$
may lead to ultraviolet singularities.

Decomposing the integral $I$ according to (\ref{decompo}), one obtains nine terms that can be grouped into six different classes -- for
simplicity the derivatives are omitted:
\begin{eqnarray}
\label{SixTypes}
A:& & G^T(x) \, G^T(x) \, G^T(-x) \, G^T(-x) ,
\nonumber \\
B:& & \Delta(x) \, G^T(x) \, G^T(-x) \, G^T(-x) \, , \ G^T(x) \, G^T(x) \, \Delta(-x) \, G^T(-x) , \nonumber \\
C:& & {\Delta}^2(x) \, G^T(-x) \, G^T(-x) \, , \
G^T(x) \, G^T(x)  \,\Delta^2(-x), \nonumber \\
D:& & \Delta(x) \, G^T(x) \, \Delta(-x) \, G^T(-x)\, ,
\nonumber \\
E:& & {\Delta}^2(x) \, \Delta(-x) \, G^T(-x) \, , \
\Delta(x) \, G^T(x) \, {\Delta}^2(-x) \, , \nonumber \\
F:& & {\Delta}^2(x) \, {\Delta}^2(-x).
\end{eqnarray}
Note that the product $\Delta(x) \Delta(-x)$ of zero-temperature propagators is proportional to $\Theta(x_4) \Theta(-x_4)$, such that
terms of the classes $D, E$ and $F$ do not contribute. Hence we are left with the cases $A, B$ and $C$. 

Classes $A$ and $B$ do not pose any problems as the corresponding integrals over the torus,
\begin{equation}
{\int}_{\! \! \! {\cal T}} \! \! d^{d_s+1}x \, \Big( {\partial}_{r} G^T {\partial}_{r}
G^T {\partial}_{s} {\tilde G}^T {\partial}_{s} {\tilde G}^T
+ 4 {\partial}_{r} \Delta \, {\partial}_{r} G^T {\partial}_{s} {\tilde G}^T
 {\partial}_{s} {\tilde G}^T \Big) \, ,
\end{equation}
are finite at $d_s$=1.

Concerning class $C$, let us consider the term
\begin{equation}
\label{termClassC}
{\partial}_{x_1} \Delta(x) \, {\partial}_{x_1} \Delta(x) \, {\partial}_{x_1} G^T(-x) \, {\partial}_{x_1} G^T(-x) \, ,
\end{equation}
where we have displayed the derivatives. In the limit $d_s \!\to \! 1$, the zero-temperature piece ${\partial}_{x_1} \Delta(x)$ can be
written as
\begin{equation}
{\partial}_{x_1} \Delta(x) \propto \frac{x_1}{{x_4}^{\frac{3}{2}}} \,
\exp\Big[ - \frac{{x_1}^2}{4 \gamma x_4} \Big] \, .
\end{equation}
On the other hand, the first term in the Taylor expansion of ${\partial}_{x_1} G^T(-x)$ at the origin, is linear in $x_1$,
\begin{equation}
\label{Taylor}
{\partial}_{x_1} G^T(-x) = {\partial}^2_{x_1} G^T(-x)|_{x=0} \, x_1 + {\cal O}({x_1}^3) \, .
\end{equation}
Accordingly,  in the limit $d_s \!\to \! 1$, the contribution in the integral $I$ takes the form,
\begin{equation}
\label{TaylorSingular}
I \, \propto \, \int \! \! d x_1 \, dx_4 \, {\Big( \frac{x_1}{{x_4}^{\frac{3}{2}}} \Big)}^2 \, e^{{-x_1}^2/{2 \gamma x_4}} \, {x_1}^2 \
\propto \ \int \! \! dx_4 \, \frac{1}{\sqrt{x_4}} \, ,
\end{equation}
which is not singular in the ultraviolet. Unlike in two or three spatial dimensions (see Refs.~\citep{Hof11a,Hof12b}), contributions of
class $C$ are finite, such that these integrals can be evaluated numerically without further ado -- in appendix \ref{appendixA} we provide
useful representations. Still, in order to check consistency of our method, in appendix \ref{appendixB} we proceed along the lines of
Ref.~\citep{GL89} and show that both variants of the method yield the same result.

Gathering all terms that contribute up to order $p^5$, and writing the integral $I$ as
\begin{equation}
I(\sigma) = T^{\frac{5}{2}} \, \frac{i(\sigma)}{\gamma^{\frac{7}{2}}} \, , \qquad \qquad
\sigma = \mu H \beta = \frac{\mu H}{T} \, , \quad \gamma = \frac{F^2}{\Sigma} \, ,
\end{equation}
the final expression for the low-temperature series of the free energy density of the ferromagnetic spin chain in a weak magnetic field is
\begin{eqnarray}
\label{FreeCollectOrder5}
z & = & - \Sigma \mu H  - \frac{1}{2 \pi^{\frac{1}{2}} \gamma^{\frac{1}{2}}} \, T^{\frac{3}{2}} \, \sum^{\infty}_{n=1}
\frac{e^{- \mu H n \beta}}{n^{\frac{3}{2}}}
- \frac{3 l_3}{4 {\pi}^{\frac{1}{2}} \Sigma {\gamma}^{\frac{5}{2}}} \, T^{\frac{5}{2}} \sum^{\infty}_{n=1}
\frac{e^{- \mu H n \beta}}{n^{\frac{5}{2}}} \nonumber \\
& & - \frac{1}{2{\Sigma}^2 {\gamma}^{\frac{3}{2}}} \, i(\sigma) \, T^{\frac{5}{2}} \; + \; {\cal O}(p^6) \, .
\end{eqnarray}
The structure of this series has been investigated previously \citep{Tak86,Tak87a,KSK03} within spin-wave theory. However, except for
Ref.~\citep{KSK03}, the authors restricted themselves to the case of {\it free} magnons. Although an interaction correction was given in
Ref.~\citep{KSK03}, this correction, as we discuss in the next section, cannot be quite correct. We emphasize that the effective
Lagrangian technique is completely systematic, unlike other approaches which are plagued with approximations or ad hoc assumptions.

\section{Ferromagnetic Spin Chains in a Magnetic Field: Thermodynamics}
\label{Thermodyn}

We now address the thermodynamic behavior of ferromagnetic spin chains, based on the representation (\ref{FreeCollectOrder5}) for the free
energy density. We first discuss the low-temperature series for the pressure that can be obtained from the temperature-dependent part of
the free energy density,
\begin{equation}
\label{Pz}
P = z_0 - z \, ,
\end{equation}
because the system is homogeneous. Up to order $p^5$, we get
\begin{equation}
\label{Pressure}
P = {\hat \eta}_0 \, T^{\frac{3}{2}} \, + \, {\hat \eta}_1 \, T^{\frac{5}{2}} \, + \, {\cal O}(p^6) \, ,
\end{equation}
with coefficients ${\hat \eta}_i$ given by
\begin{eqnarray}
\label{PressureCoefficients}
{\hat \eta}_0 & = &  \frac{1}{2 \pi^{\frac{1}{2}} \gamma^{\frac{1}{2}}} \, \sum^{\infty}_{n=1}\frac{e^{- \mu H n \beta}}{n^{\frac{3}{2}}} \, ,
\nonumber \\
{\hat \eta}_1 & = & \frac{3 l_3}{4 {\pi}^{\frac{1}{2}} \Sigma {\gamma}^{\frac{5}{2}}} \, \sum^{\infty}_{n=1}
\frac{e^{- \mu H n \beta}}{n^{\frac{5}{2}}} + \frac{1}{2{\Sigma}^2 {\gamma}^{\frac{3}{2}}} \, i(\sigma) \nonumber \\
& = & {\hat \eta}_1^{free} + {\hat \eta}_1^{int}  \, .
\end{eqnarray}
The spin-wave interaction starts manifesting itself at order $p^5 \propto T^{5/2}$ through the three-loop contribution proportional to the
dimensionless function $i(\sigma)$. The other contributions in the pressure stem from one-loop graphs, i.e., they refer to noninteracting
magnons.

\begin{figure}
\includegraphics[width=14cm]{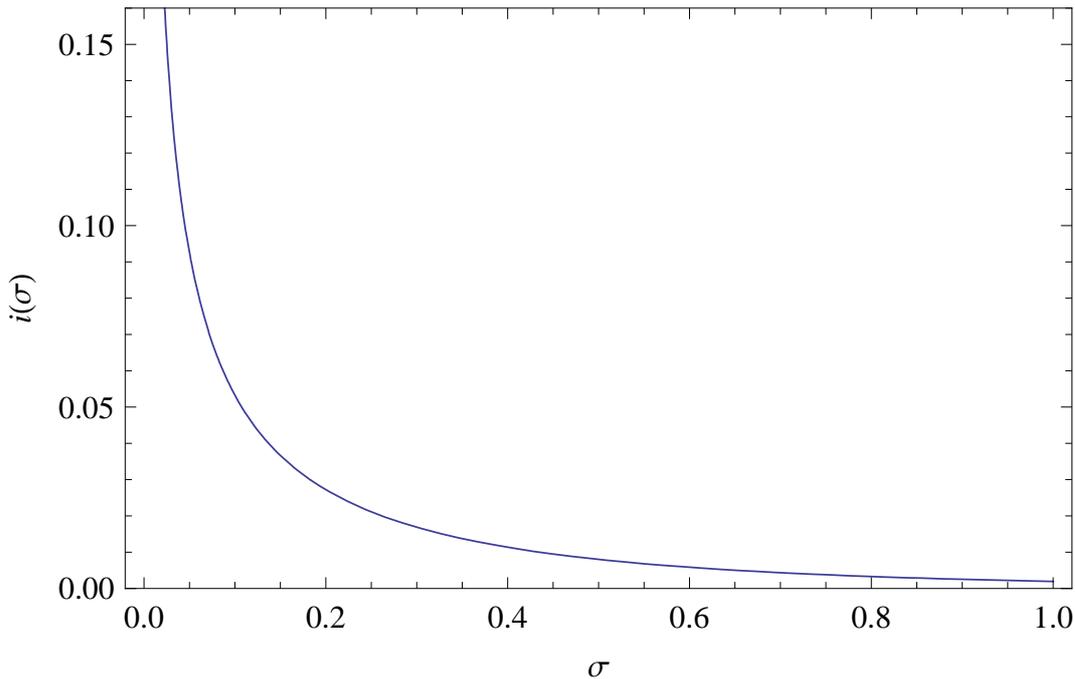}
\caption{The function $i(\sigma)$ representing the three-loop spin-wave interaction contribution in the low-temperature dynamics of the
ferromagnetic spin chain in a weak magnetic field. The quantity $\sigma$ is the dimensionless parameter $\sigma = \mu H / T$.}
\label{figure2}
\end{figure}

The spin-wave interaction is thus governed by the function $i(\sigma)$ which we have depicted in Fig.~\ref{figure2}. Since the function
$i(\sigma)$ is positive in the whole $\sigma$-range, the spin-wave interaction in the pressure always is repulsive. The smaller the ratio
between magnetic field and temperature, the stronger the repulsive interaction in the pressure gets. However, as we discuss below, for
small values of $\sigma$ the effective expansions are only valid if the temperature is extremely low.

\begin{figure}
\includegraphics[width=14cm]{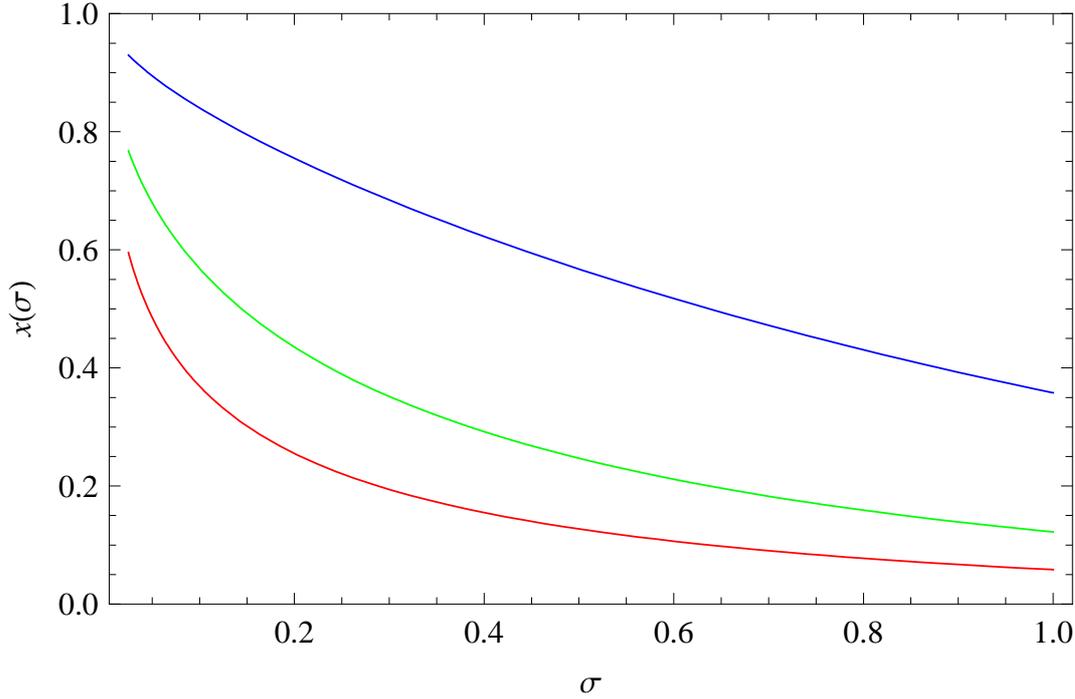}
\caption{Interaction contribution in the term of order $T^{5/2}$ in the pressure according to Eq.~(\ref{intFree}). The curves refer to
$S=\{\frac{1}{2}, 1, \frac{3}{2}\}$ from top to bottom in the figure.}
\label{figure3}
\end{figure}

Although we are dealing with a three-loop effect (next-to-next-to-leading order $T^{5/2}$ in the pressure), the effect of the interaction
is visible. This can be appreciated in Fig.~\ref{figure3}, where we have plotted the ratio
\begin{equation}
\label{intFree}
x(\sigma) = \frac{{\hat \eta}_1^{int}}{{\hat \eta}_1^{free} + {\hat \eta}_1^{int}} = { \Bigg( \frac{S^2}{16 \sqrt{\pi}}
\frac{\sum^{\infty}_{n=1} \frac{e^{- \sigma n}}{n^{5/2}}}{i(\sigma)} + 1 \Bigg)}^{-1}, \quad \sigma = \frac{\mu H}{T}
\end{equation}
for the cases  $S=\{\mbox{$\frac{1}{2}$},1,\mbox{$\frac{3}{2}$} \}$. The higher the spin $S$, the smaller the impact of the spin-wave
interaction in the pressure. Note that we have expressed the effective constants $\Sigma, \gamma$ and $l_3$ in terms of microscopic
quantities according to Eq.~(\ref{micro}).

We find it quite remarkable that the interaction contribution given in Eq.~(\ref{PressureCoefficients}) does not involve any higher-order
effective constants, but is completely determined by the zero-temperature spontaneous magnetization $\Sigma$ and the spin stiffness
$\gamma = F^2 / \Sigma$ that appear in the leading-order Lagrangian ${\cal L}^2_{eff}$. The restrictions imposed by the symmetries are thus
very strong in one spatial dimension. In particular, the fact that the spin-wave interaction in the pressure is repulsive, follows from
symmetry considerations alone.

Let us derive the low-temperature series for the energy density $u$, entropy density $s$, and heat capacity $c_V$ of the ferromagnetic
spin chain,
\begin{equation}
\label{Thermodynamics}
s = \frac{{\partial}P}{{\partial}T} \, , \qquad u = Ts - P \, , \qquad 
c_V = \frac{{\partial}u}{{\partial}T} = T \, \frac{{\partial}s}{{\partial}T} \, .
\end{equation}
Using the representation (\ref{Pressure}) for the pressure, we obtain
\begin{eqnarray}
u & = & \frac{1}{2 \pi^{\frac{1}{2}} \gamma^{\frac{1}{2}}} \, T^{\frac{3}{2}} \, \Bigg\{ \sigma \sum^{\infty}_{n=1}
\frac{e^{- \sigma n}}{n^{\frac{1}{2}}}
+ \mbox{$\frac{1}{2}$} \sum^{\infty}_{n=1} \frac{e^{- \sigma n}}{n^{\frac{3}{2}}} \Bigg\} \nonumber \\
& & + \frac{3 l_3}{4 {\pi}^{\frac{1}{2}} \Sigma {\gamma}^{\frac{5}{2}}} \, T^{\frac{5}{2}} \, \Bigg\{ \sigma \sum^{\infty}_{n=1}
\frac{e^{- \sigma n}}{n^{\frac{3}{2}}} + \mbox{$\frac{3}{2}$} \sum^{\infty}_{n=1} \frac{e^{- \sigma n}}{n^{\frac{5}{2}}} \Bigg\} \nonumber \\
& & + \frac{1}{2 {\Sigma}^2 {\gamma}^{\frac{3}{2}}} \, T^{\frac{5}{2}} \,  \Bigg\{ \mbox{$\frac{3}{2}$} i(\sigma)
- \sigma \frac{d i(\sigma)}{d \sigma} \Bigg\} + {\cal O}(p^6) \, , \nonumber
\end{eqnarray}
\begin{eqnarray}
s & = & \frac{1}{2 \pi^{\frac{1}{2}} \gamma^{\frac{1}{2}}} \, T^{\frac{1}{2}} \, \Bigg\{ \sigma \sum^{\infty}_{n=1}
\frac{e^{- \sigma n}}{n^{\frac{1}{2}}}
+ \mbox{$\frac{3}{2}$} \sum^{\infty}_{n=1} \frac{e^{- \sigma n}}{n^{\frac{3}{2}}} \Bigg\} \nonumber \\
& & + \frac{3 l_3}{4 {\pi}^{\frac{1}{2}} \Sigma {\gamma}^{\frac{5}{2}}} \, T^{\frac{3}{2}} \, \Bigg\{ \sigma \sum^{\infty}_{n=1}
\frac{e^{- \sigma n}}{n^{\frac{3}{2}}} + \mbox{$\frac{5}{2}$} \sum^{\infty}_{n=1} \frac{e^{- \sigma n}}{n^{\frac{5}{2}}} \Bigg\} \nonumber \\
& & + \frac{1}{2 {\Sigma}^2 {\gamma}^{\frac{3}{2}}} \, T^{\frac{3}{2}} \,  \Bigg\{ \mbox{$\frac{5}{2}$} i(\sigma)
- \sigma \frac{d i(\sigma)}{d \sigma} \Bigg\} + {\cal O}(p^4) \, , \nonumber \\
c_V & = & \frac{1}{2 \pi^{\frac{1}{2}} \gamma^{\frac{1}{2}}} \, T^{\frac{1}{2}} \, \Bigg\{ {\sigma}^2 \sum^{\infty}_{n=1}
\frac{e^{- \sigma n}}{n^{-\frac{1}{2}}}
+ \sigma \sum^{\infty}_{n=1} \frac{e^{- \sigma n}}{n^{\frac{1}{2}}}
+ \mbox{$\frac{3}{4}$} \sum^{\infty}_{n=1} \frac{e^{- \sigma n}}{n^{\frac{3}{2}}} \Bigg\} \nonumber \\
& & + \frac{3 l_3}{4 {\pi}^{\frac{1}{2}} \Sigma {\gamma}^{\frac{5}{2}}} \, T^{\frac{3}{2}} \, \Bigg\{ {\sigma}^2 \sum^{\infty}_{n=1}
\frac{e^{- \sigma n}}{n^{\frac{1}{2}}} + 3 \sigma \sum^{\infty}_{n=1} \frac{e^{- \sigma n}}{n^{\frac{3}{2}}} + \mbox{$\frac{15}{4}$}
\sum^{\infty}_{n=1} \frac{e^{- \sigma n}}{n^{\frac{5}{2}}} \Bigg\} \nonumber \\
& & + \frac{1}{2 {\Sigma}^2 {\gamma}^{\frac{3}{2}}} \, T^{\frac{3}{2}} \,  \Bigg\{ \mbox{$\frac{15}{4}$} i(\sigma)
- 3 \sigma \frac{d i(\sigma)}{d \sigma} + {\sigma}^2  \frac{d^2 i(\sigma)}{d {\sigma}^2} \Bigg\} + {\cal O}(p^4) \, .
\end{eqnarray}
In the above series, the contributions due to the spin-wave interaction enter at order $p^5 \propto T^{5/2}$ ($p^3 \propto T^{3/2}$) for
$u$ ($s, c_V$). In particular, there is no interaction term of order $p^4 \propto T^2$ in $u$, and no interaction term of order
$p^2 \propto T$ in $s$ and $c_V$. This is because the two-loop diagram 4 of Fig.~\ref{figure1} turns out to be 0 as a consequence of
parity \citep{Hof12c}.

Finally we turn to the magnetization and the susceptibility. With the representation (\ref{FreeCollectOrder5}) for $z$, the
low-temperature expansion for the magnetization,
\begin{equation}
\Sigma(T,H) \, = \, - \frac{\partial z}{\partial(\mu H)} \, ,
\end{equation}
amounts to
\begin{equation}
\label{magnetizationD1}
\frac{\Sigma(T,H)}{\Sigma} \; = \; 1 - {\tilde \alpha}_0 \, T^{\frac{1}{2}} - {\tilde \alpha}_1 \, T^{\frac{3}{2}} + {\cal O}(p^4) \, .
\end{equation}
The coefficients $\tilde \alpha_i$ depend on the ratio $\sigma = \mu H/T$ and read
\begin{eqnarray}
\label{SigmaCollectT(H=0)}
{\tilde \alpha}_0 & = & \frac{1}{2 {\pi}^{\frac{1}{2}} \Sigma {\gamma}^{\frac{1}{2}}} \, \sum^{\infty}_{n=1} \frac{e^{- \sigma n}}{n^{\frac{1}{2}}}
\, , \nonumber \\
{\tilde \alpha}_1 & = & \frac{3 l_3}{4 {\pi}^{\frac{1}{2}} {\Sigma}^2 {\gamma}^{\frac{5}{2}}} \, \sum^{\infty}_{n=1}
\frac{e^{- \sigma n}}{n^{\frac{3}{2}}} - \frac{1}{2 {\Sigma}^3 {\gamma}^{\frac{3}{2}}} \frac{d i(\sigma)}{d \sigma} \, .
\end{eqnarray}
On the other hand, the susceptibility of the ferromagnetic spin chain,
\begin{equation}
\chi(T,H) \, = \, \frac{\partial \Sigma(T,H)}{\partial(\mu H)} \, ,
\end{equation}
takes the form
\begin{equation}
\chi(T,H) \; = \; {\tilde \kappa}_0 \, T^{-\frac{1}{2}} + {\tilde \kappa}_1 \, T^{\frac{1}{2}} + {\cal O}(p^2) \, ,
\end{equation}
the coefficients ${\tilde \kappa}_i$ given by
\begin{eqnarray}
{\tilde \kappa}_0 & = & \frac{1}{2 {\pi}^{\frac{1}{2}} {\gamma}^{\frac{1}{2}}} \, \sum^{\infty}_{n=1} \frac{e^{- \sigma n}}{n^{-\frac{1}{2}}}
\, , \nonumber \\
{\tilde \kappa}_1 & = & \frac{3 l_3}{4 {\pi}^{\frac{1}{2}} {\Sigma} {\gamma}^{\frac{5}{2}}} \, \sum^{\infty}_{n=1}
\frac{e^{- \sigma n}}{n^{\frac{1}{2}}} + \frac{1}{2 {\Sigma}^2 {\gamma}^{\frac{3}{2}}} \frac{d^2 i(\sigma)}{d {\sigma}^2} \, .
\end{eqnarray}
Note that these series become meaningless in the limit $H \to 0$, because the leading coefficient ${\tilde \alpha}_0$ in the
magnetization, and the coefficients ${\tilde \kappa}_0$ and ${\tilde \kappa}_1$ in the susceptibility then diverge. Indeed, as we have
discussed at length in section IV of Ref.~\citep{Hof12c}, it is conceptually inconsistent to take the limit $\sigma = \mu H/T\to 0$
(temperature fixed), because we are then outside the domain where the effective expansion presented here is valid.  The point is that an
energy gap is generated nonperturbatively at finite temperatures, such that the correlation length of the magnons no longer is infinite.
This means (see section IV of Ref.~\citep{Hof12c}) that, for a given temperature, the magnetic field cannot be arbitrarily weak. Rather,
the condition
\begin{equation}
\label{constraint}
\frac{\mu H}{J} \ge 800 \frac{T^2}{J^2} \qquad (S=\mbox{$\frac{1}{2}$})
\end{equation}
must be satisfied. In this regime, the above low-temperature expansions are on safe grounds.

The restriction (\ref{constraint}) is equivalent to
\begin{equation}
\label{restriction}
\sigma \ge 800 \frac{T}{J} \qquad (S=\mbox{$\frac{1}{2}$}) \, ,
\end{equation}
implying that only at very low temperatures ($T/J \ll 1$), the parameter $\sigma$ can take small values. So while it is true that the
repulsive interaction gets stronger if we approach the limit $H \to 0$ (while keeping the temperature fixed), we have to keep in mind that
the temperature has to be very low. 

We emphasize that almost all previous investigations on ferromagnetic spin chains neglected the effect of the spin-wave interaction.
Apparently, to address the impact of the spin-wave interaction in the low-temperature series with e.g. spin-wave theory up to the order
considered here, is very challenging. Still, there is one exception, Ref.~\citep{KSK03} based on spin-wave theory at constant order
parameter, where the interaction is discussed. The explicit expression for the magnetization derived there is
\begin{eqnarray}
\label{KopietzSWT}
\frac{m(H)}{S} & = & 1 - \frac{\zeta(\mbox{$ \frac{1}{2} $})}{2 S \sqrt{\pi}} \, \sqrt{t} - \frac{1}{2 S} \, \sqrt{\frac{t}{v}}
+ \frac{1}{16} {\Bigg( \frac{1}{2 S} \sqrt{\frac{t}{v}} \Bigg)}^3 + {\cal O}(t, t^{3/2}v^{-1/2}) \, , \nonumber \\
& & t = \frac{T}{J S} \, , \quad v = \frac{H}{T} \, .
\end{eqnarray}
Up to the last term, the above expansion agrees with the leading terms of our effective result (\ref{magnetizationD1}), if we Taylor
expand in the parameter $\sigma$. However, the last term -- the interaction correction -- has no analog in the systematic effective
expansion, hinting at an inconsistency. The problem is solved by noticing that the interaction correction in (\ref{KopietzSWT}) lies
outside the parameter regime where the spin-wave picture is valid. In fact, the authors of Ref.~\citep{KSK03} say that the coefficient
$(2S)^{-1} \sqrt{t/v}$ is close to unity. For $S=\mbox{$\frac{1}{2}$}$ this means $v = t$ or, equivalently,
\begin{equation}
\frac{\mu H}{J} = 2 \frac{T^2}{J^2} \qquad (S=\mbox{$\frac{1}{2}$}) \, .
\end{equation}
So the constraint (\ref{constraint}) is not satisfied and we are clearly outside the regime where the spin-wave picture applies. Moreover,
the expansion (\ref{KopietzSWT}) appears to be inconsistent itself for the following reason: the third term in Eq.~(\ref{KopietzSWT}),
i.e., $-(2S)^{-1} \sqrt{t/v}$, is close to -1, which is not a small correction to the magnetization $m(H)/S$.

At the end of this section, we point out an important observation regarding bound states. Although the formation of two-magnon or
multi-magnon bound states becomes more important in lower space dimensions \citep{Kef66}, unfortunately, the role of bound states in the
low-temperature behavior of ferromagnetic spin chains is not well explored. In the various references cited in the present work, the
existence and the effect of bound states is not really addressed. On the other hand, a simple scaling argument \citep{Dmi12} indicates
that magnon bound states in $d_s$=1 at most start to show up at order $T^{5/2}$ in the free energy density, i.e., at the order where the
spin-wave interaction sets in.

To rigorously explore the effect of bound states, it would be interesting to compare the magnitude of the $T^{5/2}$-coefficient in our
effective expansion for the free energy density (\ref{FreeCollectOrder5}) with Bethe-ansatz results or numerical simulations. In either
case one could determine whether bound states start manifesting themselves at order $T^{5/2}$ and, if so, which part in the
$T^{5/2}$-coefficient is due to spin-waves and which part originates from the presence of magnon bound states. Unfortunately, references
based on e.g. Bethe ansatz methods, providing low-temperature expansions for the domain where our series are valid (weak but not tiny or
zero magnetic field), appear to be unavailable. Alternatively, one could also incorporate magnon bound states as explicit degrees of
freedom into the effective Lagrangian formalism and analyze their impact on the low-temperature expansions. Work in this direction is in
progress.

\section{Conclusions}
\label{Summary}

Whereas a rather considerable number of articles has been devoted to the impact of the spin-wave interaction in the three-dimensional
ideal ferromagnet over the years -- pioneered by the landmark papers by Dyson \citep{Dys56} and Zittartz \citep{Zit65} -- the analogous
question regarding ferromagnets in two spatial dimensions or ferromagnetic spin chains has been largely ignored. The present study, as
well as the preceding Refs.~\citep{Hof12a,Hof12b,Hof12c}, aimed at closing this gap in the condensed matter literature.

Within the framework of effective Lagrangians, here we have derived the partition function for the ferromagnetic spin chain up to
three-loop order and have discussed the low-temperature series for various thermodynamic quantities in a weak magnetic field, including
the magnetization and the susceptibility. In particular, we have considered the impact of the spin-wave interaction and have shown that in
the free energy density, the interaction starts manifesting itself at order $T^{5/2}$. While this power of temperature immediately follows
from the systematic loop counting, the derivation of the corresponding coefficient required the renormalization and numerical evaluation
of a specific three-loop graph which was quite elaborate. Remarkably, the coefficient of the order-$T^{5/2}$ interaction term in the
pressure is positive, such that the spin-wave interaction is repulsive.

Although various authors have also investigated the low-temperature behavior of ferromagnetic spin chains -- using, e.g., spin-wave
theory, Bethe ansatz, Schwinger-Boson mean field theory and yet other methods -- the impact of the spin-wave interaction in the
low-temperature series has been neglected; apart from Ref.~\citep{KSK03} which we have discussed in detail. Even though the system under
consideration can in principle be treated by exact methods like the Bethe ansatz, the expansions derived from these results in the
literature all refer to either a tiny or zero magnetic field. As we have outlined, this is not the domain where the effective Lagrangian
formalism operates.

With the present paper we close a series of articles devoted to the study of the low-temperature properties of ideal ferromagnets on the
basis of the effective Lagrangian method. These studies include three- and two-dimensional ideal ferromagnets as well as ferromagnetic
spin chains. In either case a complete and systematic analysis of the partition function was given up to three-loop order. By being able
to systematically go to higher orders in the perturbative expansion, as compared to conventional condensed matter techniques, we hope to
have convinced the reader that the effective Lagrangian method is indeed a very powerful tool to analyze ferromagnetic systems.

\section*{Acknowledgments}
The author would like to thank D.\ V.\ Dmitriev for correspondence.

\begin{appendix}

\section{Cateye Graph: Numerical Evaluation}
\label{appendixA}

Here we address the numerical evaluation of the three-loop graph 5c. In the relevant expression $I$,
\begin{eqnarray}
\label{I bar appendix}
I & = & {\int}_{\! \! \! {\cal T}} \! \! d^2x \, \Big( {\partial}_r G^T {\partial}_r G^T {\partial}_s {\tilde G}^T {\partial}_s
{\tilde G}^T + 4 \, {\partial}_r \Delta \, {\partial}_r G^T {\partial}_s {\tilde G}^T {\partial}_s {\tilde G}^T \Big) \nonumber \\
& & + 2 {\int}_{\! \! \! {\cal T}} \! \! d^2x \, {\partial}_r \Delta \, {\partial}_r \Delta \, {\partial}_s {\tilde G}^T
{\partial}_s {\tilde G}^T  \, ,
\end{eqnarray}
the individual terms involve the two variables $r=x_1$ and $t \! = \! x_4$, such that
\begin{equation}
d^2 x = dr \, dt \, .
\end{equation}
We introduce the dimensionless integration variables $\eta$ and $\xi$,
\begin{equation}
\eta = T x_4 \, , \qquad \xi = \frac{1}{2} \sqrt{\frac{T}{\gamma}} \, x_1 \, .
\end{equation}
In the first two terms of Eq.~(\ref{I bar appendix}), i.e. in the integrals involving quartic and triple sums, we integrate over all
space, and are thus left with one-dimensional integrals in the variable $\eta$. The expression involving the quartic sum is
\begin{eqnarray}
\label{quartic}
& & {\int}_{\! \! \! {\cal T}} \! \! d^2 x \, {\partial}_r G^T(x) \,
{\partial}_r G^T(x) \, {\partial}_s G^T(-x) \, {\partial}_s 
G^T(-x) \nonumber \\
& & = \frac{3}{32 {\pi}^{3/2} {\gamma}^{7/2}} \, T^{\frac{5}{2}} \,
{\int}^{1/2}_{\! \! \! -1/2} \! d \eta \, \sum^{\infty}_ {n_1 \dots n_4 =1} \,
e^{-{\sigma}(n_1 + n_2 + n_3 + n_4)} \ {\hat Q}(\eta,n_1,n_2,n_3,n_4) \, , \nonumber \\
& & {\hat Q}(\eta,n_1,n_2,n_3,n_4) = \frac{{\Bigg( \frac{1}{\eta + n_1}
+ \frac{1}{\eta + n_2} +\frac{1}{-\eta + n_3} +\frac{1}{-\eta + n_4}
\Bigg)}^{-5/2}}
{{\Big( (\eta + n_1)(\eta + n_2)(-\eta + n_3)(-\eta + n_4)\Big)}^{3/2}} \, ,
\end{eqnarray}
while for the triple sum we obtain
\begin{eqnarray}
\label{triple}
& & {\int}_{\! \! \! {\cal T}} \! \! d^2 x \, {\partial}_r \Delta(x) \,
{\partial}_r G^T(x) \, {\partial}_s G^T(-x) \,
{\partial}_s G^T(-x) \nonumber \\
& & = \frac{3}{32 {\pi}^{3/2} {\gamma}^{7/2}} \, T^{\frac{5}{2}} \,
{\int}^{1/2}_{\! \! \! 0} \! d \eta \, \sum^{\infty}_ {n_2 \dots n_4 =1} \,
e^{-{\sigma}(n_2 + n_3 + n_4)} \ {\hat Q}(\eta,0,n_2,n_3,n_4) \, , \nonumber \\
& & {\hat Q}(\eta,0,n_2,n_3,n_4) = \frac{{\Bigg( \frac{1}{\eta} +\frac{1}{\eta + n_2}
+ \frac{1}{-\eta + n_3} +\frac{1}{-\eta+n_4} \Bigg)}^{-5/2}}
{{\Big( \eta(\eta + n_2)(-\eta + n_3)(-\eta+n_4)\Big)}^{3/2}} \, ,
\end{eqnarray}
with
\begin{equation}
\sigma = \frac{\mu H}{T} \, , \qquad \gamma = \frac{F^2}{\Sigma} \, . 
\end{equation}
The quantities ${\hat Q}(\eta,n_1,n_2,n_3,n_4)$ and ${\hat Q}(\eta,0,n_2,n_3,n_4)$ depend in a nontrivial way on the summation variables.

The remaining integral of Eq.(\ref{I bar appendix}) which involves double sums is
\begin{eqnarray}
& & {\int}_{\! \! \! {\cal T}} \! \! d^2 x \, {\partial}_r
\Delta(x) \, {\partial}_r \Delta(x) \, {\partial}_s G^T(-x) \,
{\partial}_s G^T(-x) \nonumber \\
& & = \frac{1}{4 {\pi}^2 {\gamma}^{7/2}} \, T^{\frac{5}{2}} \,
{\int}_{\! \! \! \! 0}^{1/2} \! d \eta \, {\int}_{\! \! \! \! 0}^{\infty} \! d \xi \, {\xi}^4 \,
\sum^{\infty}_ {n_1, n_2 =1} \, e^{-{\sigma}(n_1 + n_2)} \ {\hat P}(\xi,\eta,n_1,n_2) \, ,
 \nonumber \\
& & {\hat P}(\xi,\eta,n_1,n_2) = \frac{e^{ -{\xi}^2 \Big( \frac{2}{\eta} + \frac{1}{-\eta + n_1}
+ \frac{1}{-\eta + n_2} \Big)}}
{{\Big\{ {\eta}^2(-\eta + n_1)(-\eta + n_2)\Big\}}^{3/2}} \, .
\end{eqnarray}

\section{Alternative Decomposition of the Cateye Graph}
\label{appendixB}

In one spatial dimension, contributions of class $C$ in the integral (\ref{integralI}) do not contain ultraviolet singularities and the
numerical integration over the torus can be performed directly as outlined in the preceding appendix. Still, we also want to follow the
method described in Ref.~\citep{GL89}, which provides us with an alternative decomposition of the cateye graph and therefore serves as a
consistency check.

First a sphere ${\cal S}$ of radius $|{\cal S}| \leq \beta/2$ around the origin is cut out. The integral which involves the contributions
of class $C$ is written as
\begin{eqnarray}
& & {\int}_{\! \! \! {\cal T}} \! \! d^{d_s+1} \! x \, {\partial}_r \Delta
 {\partial}_r \Delta \, {\partial}_s {\tilde G}^T {\partial}_s {\tilde G}^T
\nonumber \\
& & = {\int}_{\! \! \! {\cal S}} \! \! d^{d_s+1} \! x \, {\partial}_r \Delta
{\partial}_r \Delta \, {\partial}_s {\tilde G}^T {\partial}_s {\tilde G}^T
+ {\int}_{\! \! \! {{\cal T} \setminus \cal S}} \! \! d^{d_s+1} \! x \,
{\partial}_r \Delta {\partial}_r \Delta \, {\partial}_s {\tilde G}^T
{\partial}_s {\tilde G}^T \, .
\end{eqnarray}
In the integral over the sphere, we then subtract the term (\ref{TaylorSingular}),
\begin{eqnarray}
\label{sphereDecomp}
& & \hspace*{-1cm} {\int}_{\! \! \! {\cal S}} \! \! d^{d_s+1} \! x \, {\partial}_r \Delta(x) {\partial}_r
\Delta(x) \, {\partial}_s G^T(-x) {\partial}_s G^T(-x) \nonumber \\
& = & {\int}_{\! \! \! {\cal S}} \! \! d^{d_s+1} \! x \, {\partial}_r \Delta(x) {\partial}_r \Delta(x) \,
Q_{ss}(x) \nonumber \\
& & + {\int}_{\! \! \! {\cal S}} \! \! d^{d_s+1} \! x \, {\partial}_r \Delta(x) {\partial}_r \Delta(x) \,
{\partial}_{\alpha s} G^T(-x)|_{x=0} \, {\partial}_{\beta s} G^T(-x)|_{x=0} \, x^{\alpha} \, x^{\beta} \, ,
\end{eqnarray}
where the quantity $Q_{ss}(x)$ is given by
\begin{equation}
Q_{ss}(x) = {\partial}_s G^T(-x) {\partial}_s G^T(-x) \, - \, {\partial}_{\alpha s} G^T(-x)|_{x=0} {\partial}_{\beta s} G^T(-x)|_{x=0} \,
x^{\alpha} \, x^{\beta} \, .
\end{equation}
Finally we decompose the second integral on the right hand side of (\ref{sphereDecomp}) according to
\begin{eqnarray}
& & \hspace*{-1cm} {\int}_{\! \! \! {\cal S}} \! \! d^{d_s+1} \! x \, {\partial}_r \Delta(x) \,
{\partial}_r \Delta(x)  \,
{\partial}_{\alpha s} G^T(-x)|_{x=0} \, {\partial}_{\beta s} G^T(-x)|_{x=0} \, x^{\alpha} \, x^{\beta}
\nonumber \\ 
& = & {\int}_{\! \! \! {\cal R}} \! \! d^{d_s+1} \! x \, {\partial}_r \Delta(x) \, {\partial}_r \Delta(x)  \,
{\partial}_{\alpha s} G^T(-x)|_{x=0} \, {\partial}_{\beta s} G^T(-x)|_{x=0} \, x^{\alpha} \, x^{\beta}
\nonumber \\
& & - {\int}_{\! \! \! {\cal R} \setminus {\cal S}} \! \! d^{d_s+1} x \, {\partial}_r \Delta(x) \,
{\partial}_r \Delta(x)  \, {\partial}_{\alpha s} G^T(-x)|_{x=0} \, {\partial}_{\beta s} G^T(-x)|_{x=0} \,
x^{\alpha} \, x^{\beta} \, .
\end{eqnarray}
The integral over all Euclidean space takes the form
\begin{eqnarray}
\label{singular}
& & {\int}_{\! \! \! {\cal R}} \! \! d^{d_s+1} \! x \, {\partial}_r \Delta(x) \, {\partial}_r \Delta(x)  \,
{\partial}_{\alpha s} G^T(-x)|_{x=0} \, {\partial}_{\beta s} G^T(-x)|_{x=0} \, x^{\alpha} \, x^{\beta}
\nonumber \\
& = & \frac{d_s(d_s+2)}{2^{3d_s+5} \pi^{\frac{3d_s}{2}} \gamma^{\frac{3d_s+4}{2}}} \ T^{d_s+2} \,
{(\mu H)}^{\frac{d_s-2}{2}} \, { \Bigg\{ \sum_{n=1}^{\infty} \,
\frac{e^{- \mu H n \beta}}{n^{\frac{d_s+2}{2}}} \Bigg\} }^2 \, \Gamma(1-\frac{d_s}{2}) \, .
\end{eqnarray}
The above regularized expression is not divergent in the limit $d_s \!\to \! 1$, and reads
\begin{equation}
\frac{3}{256 \pi \gamma^{\frac{7}{2}} \sqrt{\sigma}} \, T^{\frac{5}{2}} \
{ \Bigg\{ \sum_{n=1}^{\infty} \, \frac{e^{- \sigma n}}{n^{\frac{3}{2}}} \Bigg\} }^2 \, .
\end{equation}
The integral $I$ finally amounts to
\begin{eqnarray}
\label{I bar}
I & = & {\int}_{\! \! \! {\cal T}} \! \! d^2x \, \Big( {\partial}_r G^T {\partial}_r G^T {\partial}_s {\tilde G}^T {\partial}_s
{\tilde G}^T + 4 \, {\partial}_r \Delta \, {\partial}_r G^T {\partial}_s {\tilde G}^T {\partial}_s {\tilde G}^T \Big) \nonumber \\
& & + 2 {\int}_{\! \! \! {\cal T} \setminus {\cal S}} \! \! d^2x \, {\partial}_r \Delta \, {\partial}_r \Delta \, {\partial}_s {\tilde G}^T \,
{\partial}_s {\tilde G}^T + 2 {\int}_{\! \! \! {\cal S}} \! \! d^2x \, {\partial}_r \Delta \, {\partial}_r \Delta \, Q_{ss} \nonumber \\
& & - 2 \, {\int}_{\! \! \! {\cal R} \setminus {\cal S}} \! \! d^2x \, {\partial}_r \Delta \, {\partial}_r \Delta \,
{\partial}_{\alpha s} G^T(-x)|_{x=0} \, {\partial}_{\beta s} G^T(-x)|_{x=0} \, x^{\alpha} \, x^{\beta} \nonumber \\
& & + \frac{3}{128 \pi \gamma^{\frac{7}{2}} \sqrt{\sigma}} \, T^{\frac{5}{2}} \
{ \Bigg\{ \sum_{n=1}^{\infty} \, \frac{e^{- \sigma n}}{n^{\frac{3}{2}}} \Bigg\} }^2 \, .
\end{eqnarray}
The numerical evaluation of this expression poses no problems. The following representations for the three integrals in
Eq.~(\ref{I bar}) which involve double sums have been used in our numerical evaluation:
\begin{eqnarray}
& & {\int}_{\! \! \! {{\cal T} \setminus {\cal S}}} \! \! d^2 x \, {\partial}_r
\Delta(x) \, {\partial}_r \Delta(x) \, {\partial}_s G^T(-x) \,
{\partial}_s G^T(-x) \nonumber \\
& & = \frac{1}{4 {\pi}^2 {\gamma}^{7/2}} \, T^{\frac{5}{2}} \,
{\int}_{\! \! \! \! 0}^{S} \! d \eta \, {\int}_{\! \! \! \! \sqrt{S^2-{\eta}^2}}^{\infty} \! d \xi \, {\xi}^4 \,
\sum^{\infty}_ {n_1, n_2 =1} \, e^{-{\sigma}(n_1 + n_2)} \ {\hat P}(\xi,\eta,n_1,n_2) \, ,
\nonumber \\
& & {\hat P}(\xi,\eta,n_1,n_2) = \frac{e^{ -{\xi}^2 \Big( \frac{2}{\eta} + \frac{1}{-\eta + n_1} +
\frac{1}{-\eta + n_2} \Big)}}{{\Big\{ {\eta}^2(-\eta + n_1)(-\eta + n_2)\Big\}}^{3/2}} \, ,
\end{eqnarray}
\begin{eqnarray}
& & {\int}_{\! \! \! {\cal S}} \! \! d^2 x \, {\partial}_r \Delta(x) \, {\partial}_r \Delta(x) \,
Q_{ss}(x) \\
& & = \frac{1}{4 {\pi}^2 {\gamma}^{7/2}} \, T^{\frac{5}{2}} \,
{\int}_{\! \! \! \! 0}^S \! d \eta \, {\int}_{\! \! \! \! 0}^{\sqrt{S^2-{\eta}^2}} \! d \xi \, {\xi}^4 \,
\sum^{\infty}_ {n_1, n_2 =1} \, e^{-{\sigma}(n_1 + n_2 + 2\eta)} \ {\hat Q}(\xi,\eta,n_1,n_2,\sigma) \, , \nonumber
\end{eqnarray}
with 
\begin{equation}
{\hat Q}(\xi,\eta,n_1,n_2,\sigma) = \frac{e^{ -{\xi}^2 \Big( \frac{2}{\eta} + \frac{1}{-\eta + n_1} + \frac{1}{-\eta + n_2} \Big)}
\Big[ \frac{e^{2 \eta \sigma}}{ { \{ ( -\eta + n_1)(-\eta + n_2) \} }^{3/2} }
- \frac{e^{{\xi}^2 ( \frac{1}{-\eta + n_1} + \frac{1}{-\eta + n_2} ) }}{n_1^{3/2} n_2^{3/2}}  \Big] }
{{\eta}^3}\, ,
\end{equation}
and finally,
\begin{eqnarray}
& & {\int}_{\! \! \! {\cal R} \setminus {\cal S}} \! \! d^2 x \, {\partial}_r \Delta(x) \,
{\partial}_r \Delta(x) \, {\partial}_{s \alpha} G^T(-x)|_{x=0} \,
x^{\alpha} \, {\partial}_{s \beta} G^T(-x)|_{x=0} \, x^{\beta}
\nonumber \\
& & = \ \frac{1}{4 {\pi}^2 {\gamma}^{7/2}} \, T^{\frac{5}{2}} \,
{\int}_{\! \! \! \! S}^{\infty} \! d \eta {\int}_{\! \! \! \! 0}^{\infty} \! d \xi \, {\xi}^4 \,
\sum^{\infty}_ {n_1, n_2 =1} \, e^{-\sigma(n_1 + n_2 + 2\eta)} \ {\hat R}(\xi,\eta,n_1,n_2)
\nonumber \\
& & + \ \frac{1}{4 {\pi}^2 {\gamma}^{7/2}} \, T^{\frac{5}{2}} \,
{\int}_{\! \! \! \! 0}^S \! d \eta {\int}_{\! \! \! \! \sqrt{S^2-{\eta}^2}}^{\infty} \! d \xi \, {\xi}^4 \,
\sum^{\infty}_ {n_1, n_2 =1} \, e^{-\sigma(n_1 + n_2 + 2\eta)} \ {\hat R}(\xi,\eta,n_1,n_2) \, ,
\nonumber \\
& & {\hat R}(\xi,\eta,n_1,n_2) = \frac{e^{ -2{\xi}^2/\eta}}
{{\Big\{ {\eta}^2 n_1 n_2 \Big\}}^{3/2}} \, .
\end{eqnarray}
Note that the result for the function $I$ must be independent of the radius of the sphere ${\cal S}$ -- this provides us with a welcome
consistency check on the numerics. Moreover, we have verified that the method sketched here yields the same numerical results as the one
outlined in appendix \ref{appendixA}.

\end{appendix}

\end{document}